\documentclass[french, 12pt]{article}
\usepackage[utf8]{inputenc}
\usepackage[french]{babel}
\usepackage[style=verbose]{biblatex}

\usepackage{bold-extra}

\title{ \vspace{-0.4cm} \rule{\textwidth}{1pt} \Large \textsc{Matière noire et (ou) gravitation modifiée : une approche historique et épistémologique}\\  \vspace{-0.1cm} \rule{\textwidth}{1pt} \bigskip
\textit{\large Dark matter and (or) modified gravity: a historical and \\ \vspace{-0.4cm} epistemological approach}
}
\author{\vspace{-0.5cm}\Large Benoit \textsc{Famaey} ~\& Jonathan \textsc{Freundlich}}
\date{}

\makeatletter
\renewcommand\section{\@startsection {section}{1}{\z@}%
                                   {-3.5ex \@plus -1ex \@minus -.2ex}%
                                   {2.3ex \@plus.2ex}%
                               {\normalfont\large\bfseries}}
\makeatother

\begin{document}

\maketitle

\vspace{-0.2cm}
\begin{center}
{Université de Strasbourg, CNRS, Observatoire astronomique de Strasbourg,\\ UMR 7550, F-67000 Strasbourg, France}
\end{center}

\bigskip
\vspace{1cm}

{\textbf{Résumé. -- } Le modèle standard actuel de la cosmologie suppose que la majorité de la matière dans l'Univers est constituée de matière noire et que celle-ci est fondamentalement différente de la matière ordinaire. Cette matière noire permettrait notamment d'expliquer la rotation des galaxies, le lentillage gravitationnel des amas de galaxies et l'aspect du fond diffus cosmologique, la lumière la plus ancienne de l'Univers. Mais cette matière noire existe-t-elle vraiment ? Nous revenons ici sur l'histoire de ce concept et ses implications pour la formation et l'évolution des galaxies. Nous faisons aussi le point sur les interrogations qui subsistent, les limites du modèle, et nous présentons des théories alternatives, en particulier les modifications de la gravitation qui permettraient -- peut-être -- de s'en passer.}

\medskip

{\textbf{Abstract. --} The current standard model of cosmology assumes that the majority of matter in the Universe is made of dark matter, and that the latter is fundamentally different from ordinary matter. Dark matter can in principle explain the rotation of galaxies, the gravitational lensing from galaxy clusters or the appearance of the cosmic microwave background, the oldest light in the Universe. But does dark matter really exist? Here, we review the history of this concept and its implications for the formation and evolution of galaxies. We also consider the questions that remain, the limitations of the model, and present alternative theories, in particular modifications to the gravitional law that would -- perhaps -- make it possible to do without it.}

\newpage

\section{Modéliser le mouvement des astres}


L'observation de la régularité des mouvements des astres dans le ciel suscita très tôt, probablement dès la Préhistoire, la mise en place de méthodes permettant de décrire et de prédire ces mouvements. Au moins à partir de l'Antiquité grecque, de telles méthodes purent s'appuyer sur des ``modèles", quasiment au sens moderne du terme, à savoir une représentation simplifiée, souvent formalisée mathématiquement, d'un processus ou d'un système complexe. Anaximandre de Milet (VIe siècle avant J.C.) aurait ainsi été le premier à rendre compte du mouvement des astres en imaginant un univers sphérique centré autour de la Terre. Après plusieurs siècles d'observations et de développements conceptuels, l'\textit{Almageste} de Claude Ptolémée\footnote{Cf. \url{https://ecliptiqc.ca/Almageste.php} pour une traduction en français.} (IIe siècle après J.C.) est l'aboutissement de ce modèle géocentrique et devient la référence pour décrire et prédire le mouvement des astres de l'Antiquité jusqu'à la fin du Moyen-Âge, tant en Occident que dans le monde arabe. Dans ce modèle, la Terre est immobile au centre du cosmos et le mouvement des astres est décrit grâce à des combinaisons d'orbites circulaires. En particulier, les trajectoires des planètes sont décrites grâce à un ou plusieurs épicycles, un épicycle étant un cercle dont le centre décrit lui-même un cercle, appelé déférent. Le centre du cercle déférent peut être décalé de la Terre -- il s'agit alors d'un excentrique -- et le mouvement du centre de l'épicycle ne pas être uniforme. 
Le mouvement d'une planète peut être décrit par des combinaisons différentes : par exemple, lorsque la période de révolution associée à l'épicycle est identique à celle de son centre autour de la Terre, l'épicycle peut être remplacé par un excentrique. De ce fait, on réalisa assez tôt que le mouvement du Soleil pouvait tout à fait être décrit en le supposant immobile avec la Terre en mouvement autour de lui, ce qui fut notamment envisagé par Aristarque de Samos (IIIe siècle avant J.C.). L'équivalence entre épicycle et excentrique, connue depuis Hipparque (IIe siècle avant J.C.), interroge plus généralement le statut des hypothèses proposées par les astronomes : si plusieurs combinaisons permettent de rendre compte également d'un phénomène, laquelle est conforme à la nature des choses ?

Aristote (IVe siècle avant J.C.) oppose dans sa \textit{Physique} la méthode de l'\textit{astronome mathématicien}, qui cherche à décrire, de manière abstraite, les phénomènes d'un monde céleste parfait et immuable, à celle du \textit{physicien}, qui cherche à percer la nature des corps d'un monde sublunaire désordonné\footcite{Physique}. Ptolémée et la plupart des penseurs grecs, à la suite d'Aristote, voient les hypothèses astronomiques comme des ``fictions" mathématiques, sans caractère ontologique, permettant seulement de rendre compte du mouvement des astres. Par exemple, Proclos de Lycie (Ve siècle) estime avec modestie que la raison humaine ne peut avoir qu'une image approchée des corps célestes sans jamais pouvoir en saisir l'essence, celle-ci n'étant accessible que par le divin\footcite{Procli1903procli}. Durant le Moyen-Âge, les penseurs andalous Ibn Rochd (Averroès) et Moïse Maïmonide (XIIe siècle) reprendront cette interprétation des hypothèses du modèle de Ptolémée, à savoir qu'elles ne permettent pas d'accéder à la nature des corps célestes. 
Lorsqu'est publiée en 1543 l'\oe uvre de Copernic exposant le modèle héliocentrique, \textit{De revolutionibus orbium c\oe lestium}\footcite{Copernicus1543droc.book.....C}, la préface du théologien allemand Andreas Osiander s'inscrit dans cette tradition. Cette préface précise en effet explicitement que l'héliocentrisme n'est qu'une hypothèse mathématique abstraite et que l'astronomie ne saurait percer la nature des corps célestes. 

Ce n'est que quelques dizaines d'années plus tard que cette position sera supplantée par un réalisme généralisé, voulant trouver dans les hypothèses astronomiques des affirmations sur la nature des choses, à une époque où le modèle hélio-centrique de Copernic ne permettait pas de décrire fondamentalement mieux les mouvements célestes que le modèle géocentrique de Ptolémée\footnote{Ce n'est qu'en 1838 que l'allemand Friedrich Wilhelm Bessel observera pour la première fois la parallaxe annuelle d'une étoile, reflétant le mouvement de la Terre autour du Soleil\footcite{Bessel1838AN.....16...65B}.}. Ainsi, l'astronome danois Tycho Brahé en 1578 et le jésuite allemand Christophorus Clavius en 1581 critiquent tous deux le modèle de Copernic en exigeant que les hypothèses astronomiques soient au préalable compatibles avec les principes de la physique, qui portent sur la nature des choses et ne cherchent pas juste à rendre compte mathématiquement des phénomènes observés, et les Écritures saintes, qui placent la Terre immobile au centre du cosmos. Le philosophe italien Giordano Bruno, l'astronome allemand Johannes Kepler puis Galilée font preuve du même réalisme, cette fois en affirmant que seules les hypothèses de Copernic sont conformes à la réalité et en plaçant ainsi désormais l'astronomie au sein de la physique. Ce réalisme n'était pas justifié, dans la mesure où ils transformaient une hypothèse en certitude, mais il est possible d'y voir comme Pierre Duhem\footcite{Duhem2003-DUHSLA} l'intuition qu'il n'y avait pas de distinction fondamentale entre le monde sublunaire et le monde céleste, intuition qui allait prendre tout son sens avec la théorie de la gravitation d'Isaac Newton.

Dans ses \textit{Philosophiæ naturalis principia mathematica}, en 1687, Newton énonce des principes mathématiques qui permettent de décrire le mouvement tant des corps du monde sublunaire que ceux du monde supralunaire\footcite{Newton}. La chute des corps sur Terre et le mouvement des planètes autour du Soleil y sont expliqués par la même hypothèse, à savoir la force universelle de la gravitation, qui caractérise l'attraction entre deux masses et dont Newton donne l'expression mathématique. Les principes du mouvement de Newton peuvent être vus comme une définition de ce qu'est une force, concept abstrait qui permet de modéliser ce qui déforme un corps ou en modifie le mouvement. Notamment parce que sa théorie implique que la force gravitationnelle est une action immédiate à distance, Newton ne s'engage pas quant à une quelconque interprétation ontologique, se plaçant plutôt dans la lignée des Anciens et d'une interprétation de la force comme une  ``fiction" mathématique. 


Le principe fondamental de la dynamique (la deuxième loi de Newton) stipule que la résultante des forces divisée par la masse d'un objet est égale à l'accélération de cet objet : les changements qui arrivent dans le mouvement sont proportionnels à la force et inversement proportionnels à la masse. L'expression mathématique de la force gravitationnelle permet de retrouver les lois empiriques décrivant les orbites elliptiques des planètes autour du Soleil, formalisées notamment par Kepler. L'ellipticité de ces orbites est faible : dans l'approximation d'une orbite circulaire uniforme et en négligeant la masse des planètes en comparaison de celle du Soleil, le carré de la vitesse $V$ d'une planète est directement proportionnel à la masse centrale $M_\odot$ du Soleil, et inversement proportionnel au rayon $r$ de l'orbite : 
\begin{equation}
V^2 = \frac{G M_\odot}{r}, 
\end{equation}
où $G$ est la constante de gravitation universelle de Newton. En connaissant la distance entre le Soleil et la Terre, la période de révolution de la Terre autour du Soleil et la valeur de la constante $G$, cette relation permet de déterminer la masse du Soleil. 

L'une des conséquences les plus spectaculaires du modèle de Newton fut la découverte de Neptune par Urbain Le Verrier en 1846. La trajectoire de la planète Uranus (découverte en 1781 grâce à leur télescope par William et Caroline Herschel) présentait des anomalies, à savoir qu'elle semblait être accélérée sur une partie de son orbite par rapport à ce qui était attendu de l'attraction du Soleil, en tenant compte de l'attraction de toutes les autres planètes, et ralentie sur une autre partie de son orbite. L'ajout d'une nouvelle planète permettait d'expliquer ces anomalies, et cette nouvelle planète suggérée par le calcul de Le Verrier fut rapidement observée là où elle était attendue et nommée Neptune. Le mouvement d'Uranus avait permis de découvrir une masse auparavant {\it invisible} grâce au modèle de Newton !
Aujourd'hui, on détecte de la même façon des planètes en dehors du Système Solaire grâce aux légères variations périodiques de la vitesse de l'étoile autour de laquelle elles tournent, et c'est aussi grâce à des mesures précises du mouvement des étoiles au centre de la Voie Lactée que fut mis en évidence le trou noir qui s'y trouve\footcite{Ghez2008ApJ...689.1044G}\footcite{Genzel2010RvMP...82.3121G}. Qu'il s'agisse de Neptune, de ces planètes en dehors du Système Solaire, ou du trou noir central de notre Galaxie, c'est une mesure de vitesse qui permet de déterminer la masse. 

Il y a toutefois des cas où les anomalies détectées dans le mouvement des astres par rapport aux prédictions théoriques n'indiquent pas la présence d'une masse encore inconnue mais plutôt la nécessité de remettre en question, hors d'un certain domaine de validité approximative, le modèle à partir duquel sont faites les prédictions. Ainsi, les observations de Brahé permirent à Kepler de montrer que les orbites des planètes n'étaient pas circulaires comme dans le modèle de Copernic, mais elliptiques. Porté par le succès de la détection de Neptune, Le Verrier tenta quant à lui d'expliquer une anomalie dans le mouvement de Mercure par une planète hypothétique encore plus proche du Soleil, qui aurait été nommée Vulcain. La planète fut cherchée sans succès pendant plusieurs décennies, et il n'est aujourd'hui plus nécessaire de l'invoquer pour rendre compte de l'anomalie dans le mouvement de Mercure. En effet, l'interaction gravitationnelle est décrite depuis 1916 par la théorie de la relativité générale d'Albert Einstein\footcite{Einstein1915SPAW.......844E}\footcite{Einstein1916AnP...354..769E} et celle-ci explique entièrement le mouvement de Mercure : il n'y a plus d'anomalie. La théorie de la relativité générale ne fut pas introduite pour résoudre l'anomalie du mouvement de Mercure, mais pour décrire la gravitation d'une manière qui non seulement rendrait compte du \textit{principe d'équivalence}, c'est-à-dire l'égalité entre la masse inertielle du principe fondamental de la dynamique et la masse grave de la loi de la gravitation universelle, mais serait encore compatible avec la théorie de la relativité restreinte d'Einstein\footcite{Einstein1905AnP...322..891E}, c'est-à-dire avec le principe issu de l'expérience selon lequel aucune information ne peut se propager plus vite que la vitesse de la lumière. La théorie Newtonienne de la gravitation n'était pas compatible avec ce dernier principe, dans la mesure où elle supposait une action immédiate à distance. 

Dans le cadre de la relativité générale, on ne décrit plus la gravitation comme une force, mais comme une propriété intrinsèque de l'espace-temps, qui se courbe en présence de masses. Cette description s'appuie sur le  principe d'équivalence : l'accélération gravitationnelle qui s'exerce sur un corps ne dépend pas de sa masse et rien ne semble distinguer un champ de gravitation d'une accélération pour l'observateur qui se trouve dans un tel champ ou subit une accélération. La théorie de la relativité générale se ramène approximativement à la théorie de Newton lorsqu'il s'agit de décrire l'attraction gravitationnelle qu'on ressent sur Terre et le mouvement de la plupart des planètes autour du Soleil, lorsque le champ gravitationnel n'est pas aussi fort que celui dans lequel est plongé Mercure, la planète la plus proche du Soleil. 
La théorie de la relativité générale n'a pas encore été mise en défaut dans des systèmes de la taille du Système Solaire. 
En particulier, les étoiles qui s'approchent au plus près du trou noir central de la Voie Lactée, là où son attraction gravitationnelle est la plus forte, suivent très précisément les prédictions de la théorie de la relativité générale \footcite{GRAVITY2018A&A...615L..15G}. 

La gravitation est la plus faible des interactions fondamentales, ce dont on peut se rendre compte en considérant qu'il faut la masse de la Terre entière pour nous maintenir à sa surface et qu'on peut encore s'en affranchir quelques instants en sautant. Malgré cela, c'est l'interaction qui régit la dynamique de l'Univers à grande échelle, parce que les objets massifs ont toujours tendance à s'attirer, contrairement, par exemple, à des objets chargés électriquement qui tendent vers la neutralité électrique à grande échelle. C'est donc la théorie de la gravitation, la relativité générale, qui doit permettre d'expliquer les mouvements des étoiles dans les galaxies, des galaxies dans les amas de galaxies, ou la dynamique de l'Univers lui-même. Cependant, lorsqu'on applique cette théorie à la dynamique des étoiles et du gaz au sein des galaxies, et plus généralement à l'Univers à grande échelle, on est confronté, comme nous allons maintenant le voir, à un problème de masse et d'énergie manquante, qui pourrait être réminiscent des anomalies rencontrées au XIXe siècle lorsqu'il s'agissait de rendre compte des orbites d'Uranus ou de Mercure.

\medskip

\section{Le problème de la masse manquante : la matière noire }


Loin des lumières des villes modernes, les nuits claires nous laissent voir une bande blanchâtre dans le ciel : la Voie Lactée. Il fallut attendre la lunette astronomique de Galilée en 1610 pour réaliser qu'il s'agissait en fait d'une multitude d'étoiles, dont le Soleil et les étoiles visibles à l'\oe il nu ne constituent qu'une infime partie. Bien que différents penseurs et philosophes, dont Démocrite, Giordano Bruno et Emmanuel Kant, aient suggéré que l'Univers puisse comprendre une infinité de mondes analogues au nôtre, ce n'est qu'au début du XXe siècle qu'on réalisa empiriquement que l'Univers s'étendait au-delà de la Voie Lactée et qu'il contenait des centaines de milliards d'autres ``univers-îles" comme la Voie Lactée : les galaxies. Les étoiles, le gaz et la poussière qui forment une galaxie sont liés par la gravitation, et les galaxies elles-mêmes peuvent former des amas de galaxies liés par la gravitation, traçant à grande échelle une immense toile cosmique. 
Les galaxies furent initialement observées comme des ``nébuleuses" plus diffuses que les étoiles, dont il n'était pas clair qu'elles appartinssent à la Voie Lactée ou non : c'est en mesurant la distance d'une étoile individuelle dans la ``nébuleuse'' d'Andromède en 1923 qu'Edwin Hubble conclut que celle-ci se trouvait bien au-delà de la Voie Lactée et était en fait une galaxie analogue à la notre\footcite{Hubble1925PA.....33..252H}. 
La méthode de mesure de la distance de cette étoile se basait sur le fait qu'il s'agissait d'une étoile à la luminosité variable, dite {\it céphéide}, pouvant être utilisée comme une ``chandelle standard'' : la luminosité intrinsèque de ce type d'étoiles peut être connue grâce à la relation période-luminosité, découverte par l'astronome américaine Henrietta Leavitt en 1908, liant la luminosité de ces étoiles à leur variation périodique (plus leur luminosité est grande, plus la période associée est grande aussi), puis leur distance déduite de la différence entre cette luminosité intrinsèque et leur luminosité apparente\footcite{Leavitt1908AnHar..60...87L}.


Entretemps, le physicien russe Alexander Friedmann avait montré dès 1922 que l'Univers pouvait être en expansion avec un taux d'expansion calculable grâce aux équations de la relativité générale\footcite{Friedmann1922ZPhy...10..377F}, et le chanoine belge Georges Lemaître trouva indépendamment une solution similaire\footcite{Lemaitre1927ASSB...47...49L} en 1927. Hubble combina de son côté ses mesures de distance avec les mesures de Vesto Slipher et Milton Humason des décalages vers le rouge\footnote{Tout comme le son d'une ambulance qui s'éloigne de nous est plus grave qu'au repos, l'effet Doppler fait que le spectre d'une étoile ou d'une galaxie qui s'éloigne de nous est décalé vers le rouge, et ce d'autant plus qu'elle s'éloigne rapidement de nous. D'un point de vue cosmologique, on peut aussi considérer que l'expansion de l'Univers étire la longueur d'onde de la lumière s'y propageant.} du spectre des galaxies, montrant ainsi que l'Univers était bel et bien en expansion\footcite{Hubble1929PNAS...15..168H}. Cette expansion peut être localement caractérisée par une constante, appelée la constante de Hubble-Lema\^itre, dont Hubble estima observationnellement la valeur à 500 (km/s)/Mpc. Cette constante, qui représente la vitesse à laquelle s'éloigne une galaxie située à 1 Mpc\footnote{1 Mpc = $10^6$ parsecs, 1 parsec correspondant à 3,26 années-lumière.\label{footnote:Mpc}} de l'observateur du fait de l'expansion de l'Univers, est actuellement estimée à environ sept fois moins\footcite{Riess2022ApJ...934L...7R}, mais les observations indiquent bel et bien un Univers en expansion. Toutefois, la valeur exacte de cette constante est aujourd'hui sujette à controverse, car la valeur estimée dans le cadre du modèle cosmologique standard ne correspond pas exactement à la valeur mesurée plus directement à partir de la corrélation entre distance et décalage vers le rouge. Il s'agit là d'un des plus grands mystères actuels de la cosmologie.

L'astronome suisse Fritz Zwicky fut l'un des premiers à s'intéresser à la dynamique des structures extra-galactiques que sont les amas de galaxies. Il mesura en 1933 la dispersion de vitesse le long de la ligne de visée de galaxies de l'amas de Coma, grâce à l'effet Doppler, et en conclut que la masse totale de l'amas était des centaines de fois plus importante que la masse visible en son sein\footcite{Zwicky1933AcHPh...6..110Z}. Il nomma la masse additionnelle nécessaire {\it Dunkle Materie}, ce qui fut traduit par ``matière noire" en français. La dispersion des vitesses qu'il mesura avec un nombre limité de galaxies de l'amas était très proche de la valeur mesurée aujourd'hui et, même si la valeur du rapport entre la masse de l'amas déduite de cette dispersion et la masse de matière ordinaire était surestimée (notamment parce qu'il utilisait la grande valeur de la constante de Hubble-Lema\^itre de 500 (km/s)/Mpc, sous-estimant la distance de l'amas et la luminosité des galaxies y résidant, et parce qu'il n'avait pas accès à la masse de gaz chaud dans l'amas, celle-ci émettant en rayons X), ce problème de masse manquante dans les amas de galaxies n'a toujours pas disparu : le rapport entre masse totale et masse de matière ordinaire est aujourd'hui estimé être de l'ordre de 6. Ce problème ne fut toutefois pas particulièrement pris au sérieux pendant plusieurs décennies. 


La possibilité de la présence de grandes quantités de matière invisible aux échelles galactiques et au-delà revint sur le devant de la scène une quarantaine d'années plus tard, tout d'abord via des considérations purement théoriques. En effet, les premières simulations numériques de disques galactiques en rotation n'étant pas stables, les chercheurs américains Jerry Ostriker et Jim Peebles proposèrent en 1973 que ces disques soient entourés de grands halos sphériques de matière invisible permettant de les stabiliser\footcite{Ostriker1973ApJ...186..467O} : de la matière noire. Même si on se rendit compte une dizaine d'années plus tard que cet argument théorique n'était pas tout à fait correct, l'hypothèse d'Ostriker et Peebles fut généralement acceptée dès la fin des années 1970 et corroborée observationnellement par le fait que les vitesses de rotation mesurées à très grande distance du centre des galaxies étaient constantes, phénomène observé par différents groupes mais en particulier plus systématiquement par les équipes de Vera Rubin et d'Albert Bosma\footcite{Rogstad1974}\footcite{Roberts1975}\footcite{Rubin1978ApJ...225L.107R}\footcite{Bosma1978PhDT.......195B}\footcite{Faber1979}\footnote{Pour un compte-rendu détaillé, nous renvoyons le lecteur à l'article d'Albert Bosma dans cette publication collective.}. 

La grande majorité de la masse visible, dite ``baryonique"\footnote{Les protons et les neutrons, dont la mass  domine largement la masse de matière ``ordinaire" dans l'Univers, sont des baryons ; par abus de langage, toute la matière ``ordinaire'' est souvent appelée matière ``baryonique".}, $M_b$, d’une galaxie se situe dans les quelques kiloparsecs\footnote{1 kpc (kiloparsec) = $10^3$ parsecs, soit à peu près 3 261 années-lumière, cf.  note \ref{footnote:Mpc}.} centraux ; dans les régions externes, le carré de la vitesse $V$ du gaz sur des orbites quasi-circulaires à une distance $R$ du centre galactique devrait donc être proportionnel à $M_b$ et inversement proportionnel à $R$, comme dans l'équation~(1) pour le Système Solaire, avec    
\begin{equation}
V^2 \simeq \frac{GM_b}{R}. 
\end{equation} 
Or la vitesse circulaire, mesurée grâce à l'effet Doppler subi par l'émission de l'hydrogène atomique de part en part de la galaxie en rotation, ne décroît pas comme la racine carrée de la distance et a plutôt tendance à rester constante à très grandes distances du centre, son carré suivant\footcite{McGaugh2000ApJ...533L..99M}\footcite{Lelli2016ApJ...816L..14L}
\begin{equation}
V^2 \propto \sqrt{M_b}. 
\end{equation} 
Le même type d'écart est par la suite apparu pour la température du gaz des amas de galaxies, trop chaude, et pour l’angle de déviation des rayons lumineux autour des galaxies et des amas de galaxies, trop important, impliquant de nouveau des masses plus grandes que ce que l'on observe. En effet, le gaz chaud des amas ne peut être retenu dans l'amas que si la masse de ce dernier est importante, et la masse des galaxies et des amas agit comme une ``lentille gravitationnelle'' qui courbe les trajectoires des rayons lumineux. 
Aujourd'hui, on a besoin de la matière noire pour expliquer toute une série d'observations allant de l'échelle des galaxies jusqu'au fond diffus cosmologique, la lumière la plus ancienne de l'Univers, émise il y a un peu moins de 13,7 milliards d'années. 

L'introduction d'une masse invisible pour expliquer la vitesse de rotation du gaz dans les galaxies à la fin des années 1970 rappelle par certains aspects la résolution de l'anomalie de l'orbite d'Uranus.  Mais il faut aussi se souvenir de la façon dont fut résolue l'anomalie de l'orbite de Mercure. 
Très rapidement, les scientifiques se sont donc demandé si cette matière noire pouvait être le signe des limites du domaine de validité de la relativité générale, comme ce fut le cas des anomalies du mouvement de Mercure pour la théorie de la gravitation de Newton, plutôt qu'une présence de matière invisible, comme dans le cas d'Uranus. C'est ainsi que le physicien israélien Moti Milgrom proposa dès 1983 une modification de la dynamique dans le régime des très faibles accélérations qui pouvait permettre de se passer de matière noire dans les galaxies\footcite{Milgrom1983ApJ...270..365M}\footcite{Milgrom1983ApJ...270..371M}\footcite{Milgrom1983ApJ...270..384M}. Avec une telle modification, c'est la matière noire elle-même qui serait reléguée au rang de ``fiction'' permettant de sauver les phénomènes. Pour pouvoir voir dans la matière noire une hypothèse sur la {\it nature} des choses, une détection plus directe que par son action gravitationnelle semblait nécessaire, à l'image de la découverte de Neptune. 
%
%
%
On s'est donc prestement mis en chasse de la matière noire, en se demandant notamment si elle ne pouvait pas être constituée d'objets célestes compacts ou de gaz moléculaire. Mais les développements de la cosmologie portèrent un coup fatal à ces deux possibilités, en montrant que la matière noire ne pouvait pas être constituée de matière ordinaire.

La cosmologie, en tant qu'étude de l'Univers dans son ensemble, avait en effet continué à se développer après les travaux pionniers de Friedmann et Lema\^itre, et engendré de nombreux débats concernant l'expansion de l'Univers, opposant notamment les tenants d'un Univers en expansion et ceux d'un Univers statique. Ces débats pouvaient à certains égards rappeler ceux de la Renaissance entre les tenants du modèle géocentrique et ceux du modèle héliocentrique. Les chercheurs Ralph Alpher et Robert Herman prédirent théoriquement en 1948 qu'un Univers en expansion impliquait un rayonement relique de l'Univers primitif, c'est-à-dire un rayonnement émis très tôt dans l'histoire de l'Univers dans lequel on baignerait aujourd'hui\footcite{Alpher1948Natur.162..774A}. 
Au cours de la décennie suivante, George Gamow et ses collaborateurs firent différentes estimations des caractéristiques attendues de ce rayonnement, qui pouvait être la signature indubitable d'un Univers en expansion au-delà de la relation d'Hubble-Lema\^itre. 
Mais ce sont Arno Penzias et Robert Wilson, deux employés des Bell Labs aux Etats-Unis, qui le découvrirent en 1965\footcite{Penzias1965ApJ...142..419P}. Ils utilisaient une antenne initialement destinée à interagir avec des satellites de télécommunications pour mesurer l'émission du reste de supernova Cassiopée A. 
Lorsqu'ils détectèrent une émission uniforme dans le ciel, ils se rendirent compte qu'il s'agissait du rayonnement relique, au sujet duquel Jim Peebles avait récemment donné un exposé et qui devait faire comme un ``bruit de fond'' cosmologique. Peebles collaborait à cette époque avec Robert Dicke, Peter Roll et David Wilkinson dans un effort concerté pour détecter ce rayonnement. En prenant contact avec eux, Penzias et Wilson eurent la confirmation qu'il s'agissait bel et bien de ce fond diffus\footcite{Dicke1965ApJ...142..414D}, la relique lumineuse venant du fond des âges prédite par les modèles d'univers en expansion. 
Dès sa découverte, on comprit que des fluctuations de longueur d'onde de ce rayonnement devraient correspondre à des fluctuations de température du plasma primordial et donc aux perturbations initiales ayant donné naissance aux structures de l'Univers, comme les galaxies et les amas de galaxies. 
Mais les observateurs de l'époque furent surpris de ne détecter aucune anisotropie dans le fond diffus cosmologique, c'est-à-dire aucune différence en fonction de la direction : avec des fluctuations d'amplitude inférieure à un centième, comme ils pouvaient le déduire avec leurs instruments, il semblait compliqué pour la seule matière baryonique d'évoluer vers les structures visibles aujourd'hui dans l'Univers étant donné l'âge de l'Univers. Pour expliquer la croissance de ces structures, il fallait à nouveau évoquer une matière noire additionnelle et, qui plus est, une matière noire non baryonique \footcite{Blumenthal1984} ! 

C'est seulement en 1992, avec la mission spatiale COBE, que les fluctuations anisotropes du fond diffus cosmologique furent détectées, avec une amplitude de seulement un cent-millième\footcite{Smoot1992ApJ...396L...1S}. Outre le problème de la croissance des structures, cette amplitude particulièrement faible et donc la quasi-isotropie du fond diffus cosmologique posait un autre problème fondamental : les différentes portions de ciel correspondant à l'échelle des fluctuations étaient en principe des portions d'Univers qui n'avaient pas pu avoir le temps de communiquer entre elles depuis le {\it Big Bang}, l'instant initial des modèles d'expansion de l'Univers, sauf à supposer une communication à une vitesse supérieure à la vitesse de la lumière, interdite par la relativité restreinte. Comment dès lors expliquer pareille isotropie ? La solution fut d'envisager dès le début des années 1980 une phase d'{\it inflation} initiale\footcite{Guth1981PhRvD..23..347G}, correspondant à une phase d'expansion accélérée de l'Univers, avec une valeur phénoménale de la constante de Hubble-Lema\^itre, de l'ordre de $10^{16}$ (km/s)/Mpc. À la fin de cette phase d'inflation (dont la durée minimale est fixée à $10^{-36}$ secondes mais qui pourrait avoir duré beaucoup plus longtemps, voire éternellement\footcite{Aguirre2003}), le champ (ou les champs) responsable de cette expansion accélérée, appelé l'``inflaton", se serait alors désintégré en radiation et en matière, qui seraient donc apparues seulement à cet instant. 
Les différentes portions d'Univers déconnectées au moment de l'émission du fond diffus auraient ainsi pu communiquer auparavant, c'est-à-dire être initialement connectées causalement pendant la phase d'inflation. Dans le cadre de cette théorie de l'inflation, il est naturel que la géométrie de l'Univers soit Euclidienne (univers plat), car l'inflation tend à effacer toute courbure initiale, et donc que la densité totale de l'Univers corresponde à la densité {\it critique} d'énergie pour laquelle un univers est plat d'après la relativité générale. La densité baryonique dans l'Univers, quant à elle, a une influence directe sur les rapports d'abondance initiaux entre hydrogène, deuterium, helium et lithium produits par la nucléosynthèse primordiale, processus établi dans les années 1940 par Ralph Alpher, George Gamow et Hans Bethe\footcite{Alpher1948PhRv...73..803A}. Ces rapports d'abondance peuvent être mesurés dans les spectres des quasars\footcite{Cyburt2016RvMP...88a5004C}, des galaxies primordiales particulièrement lumineuses, indiquant une densité baryonique typique de moins de 5\% de la densité critique. Ce serait donc la présence de quantités importantes de matière noire qui permettrait à l'Univers dans son ensemble d'atteindre la densité critique caractéristique d'un univers plat. Ces considérations menèrent à la fin des années 1980 à un premier modèle ``standard" de la cosmologie \footcite{Davis1988lsmu.book..439D} dans lequel la densité totale de matière dans l'Univers était la densité critique, et la fraction de matière baryonique seulement de l'ordre de quelques pourcents. 

Cependant, ce premier modèle standard fut rapidement confronté à de nouveaux problèmes. Par exemple, la fraction de matière baryonique observée dans les amas de galaxies était beaucoup plus importante que les quelques pourcents attendus. 
Où était toute cette matière noire manquante ? Ceci amena Jerry Ostriker et Paul Steinhardt à proposer en 1995 un nouveau modèle standard de la cosmologie\footcite{Ostriker1995Natur.377..600O}, dominé à notre époque (mais pas dans l'Univers primordial) par une constante cosmologique telle que proposée par Einstein en 1917 pour généraliser les équations de la relativité générale\footcite{Einstein1917SPAW.......142E}. Ainsi, la fraction de matière baryonique pouvait être de seulement quelques pourcents de l'{\it énergie} totale actuelle, dominée par la constante cosmologique, mais bien plus importante que dans le modèle précédent en termes de fraction de {\it matière}. Un tel modèle avec constante cosmologique impliquait une accélération de l'expansion de l'Univers, prédiction qui fut confirmée en 1998 par l'observation de supernovae lointaines utilisées comme chandelles standards, dont la luminosité est connue, ce qui permet une mesure de leur distance\footcite{Riess1998AJ....116.1009R}\footcite{Perlmutter1999ApJ...517..565P}\footnote{Cf. aussi les céphéides\footcite{Leavitt1908AnHar..60...87L}.}. La constante cosmologique, dont l'origine et la valeur restent encore mystérieuses aujourd'hui\footcite{Weinberg1989RvMP...61....1W}, peut être assimilée à un fluide exotique exercant une pression négative et engendrant une ``poussée" gravitationnelle expliquant l'accélération de l'expansion. Ce fluide mystérieux, qui de plus ne se dilue pas dans l'expansion de l'espace, est souvent appelé l'{\it énergie noire} et est supposé n'avoir strictement aucun lien avec la matière noire.

Ce deuxième modèle cosmologique standard, appelé modèle $\Lambda$CDM ($\Lambda$ pour la constante cosmologique et CDM pour ``cold dark matter", la matière noire ``froide", c'est-à-dire faite de particules se déplaçant à des vitesses très inférieures à celles de la lumière), a été progressivement affiné à partir des observations des grandes structures de l'Univers et surtout des fluctuations de température du fond diffus cosmologique.
Après COBE, les missions spatiales WMAP et Planck ont en effet permis d'améliorer la mesure de ces fluctuations en fonction de l'échelle angulaire sur le ciel et de produire un ``spectre de puissance angulaire" du fond diffus, cartographiant les oscillations accoustiques (c'est-à-dire des ondes de ``compression-raréfaction") du plasma primordial à l'origine des fluctuations de température\footcite{Spergel2003ApJS..148..175S}\footcite{Planck2020A&A...641A...6P}. Dans ce spectre, les différents pics correspondent à des perturbations soit compressives (les pics impairs, en commençant par le premier, celui à la plus grande échelle) soit de raréfaction (les pics pairs, comme le deuxième) du plasma. Le rapport d'amplitude des trois premiers pics permet d'estimer la densité baryonique et la densité de matière dans l'Univers primordial, dans la mesure où les pics compressifs sont amplifiés lorsque la densité baryonique augmente, tandis que les pics de raréfaction sont amplifiés dans l'ère dominée par la radiation. Or, les deuxième et troisième pics ont une amplitude très proche, ce qui n'aurait pas pu se produire si la perturbation correspondant au deuxième pic était ``entrée dans l'horizon'' 
\footnote{Après l'ère d'inflation, les zones de l'Univers séparées d'une distance plus grande que celle que peut parcourir la lumière dans le temps imparti depuis la fin de la phase d'inflation ne peuvent communiquer entre elles : elles sont séparées par un ``horizon" indépassable. Le temps s'écoulant, l'horizon s'agrandit et des perturbations de taille de plus en plus grandes peuvent ``entrer dans l'horizon" : les perturbations aux petites échelles sont entrées plus tôt dans l'horizon, tandis que les plus grandes (le premier pic) y sont entrées en dernier. Le bilan énergétique entre radiation et matière change au cours du temps, la radiation dominant initialement : les perturbations aux petites échelles sont donc entrées dans l'horizon dans ``l'ère de radiation". Sans matière noire, ce serait également le cas de la perturbation correspondant au deuxième pic, mais avec matière noire, celle-ci entre dans l'horizon dans l'ère de matière, ce qui modifie le spectre observé.} 
dans une ère dont le bilan énergétique était dominé par la densité d'énergie des photons plutôt que par la densité de la matière, car le deuxième pic de raréfaction aurait alors été amplifié. Pour résoudre ce problème, il faut une densité de matière plus grande, mais il faut ajouter de la matière non-baryonique pour ne pas amplifier indûment le premier pic : la matière noire, une fois de plus ! A partir de cette analyse détaillée du fond diffus, on peut extrapoler la quantité d'énergie contenue dans les trois grandes composantes de l'Univers aujourd'hui : 69\% d'énergie noire, 26\% de matière noire et 5\% de matière baryonique. Il s'agit donc d'un modèle qui paramétrise avec grande précision notre ignorance de 95\% de l'Univers. 

L'accord du modèle $\Lambda$CDM avec la plupart des observables cosmologiques à grande échelle est néanmoins remarquable. Par exemple, ce modèle prédisait que les oscillations accoustiques dans le plasma primordial devaient avoir laissé une empreinte dans la distribution des galaxies, plus précisément sous la forme d'une signature dans la fonction de corrélation à deux points. En 2005, ces oscillations accoustiques furent détectées exactement telles que prédites\footcite{Eisenstein2005ApJ...633..560E}. Cela illustre à quel point cette paramétrisation, si elle n'est pas pleinement satisfaisante d'un point de vue {\it ontologique}, permet une représentation fidèle des phénomènes observés à grande échelle dans l'Univers. Il est néanmoins prématuré d'adopter l'attitude de l'astronome physicien plutôt que celle de l'astronome mathématicien. En effet, le modèle géocentrique de Ptolémée était lui aussi capable de prédictions en parfait accord avec les observations. Une détection indépendante de la matière noire, via une autre interaction que la gravitation sur laquelle on se repose pour en inférer la présence, serait nécessaire pour que l'attitude de l'astronome physicien et le caractère ontologique de la matière noire puissent être validés.


\section{\`A la recherche de la matière noire}

Comme expliqué dans la section précédente, on se demanda initialement si la matière noire pouvait être baryonique, par exemple sous la forme d'objets celestes compacts ou de gaz moléculaire, mais la cosmologie a porté un coup fatal à cette hypothèse : la matière noire doit être non-baryonique. Que peut-être cette matière non-baryonique représentant près de 85\% de la matière dans l'Univers ?

Le problème de la matière noire coïncida dans la deuxième moitié du XXe siècle avec les évolutions théoriques de la physique des particules. Les travaux de Sheldon Glashow, Abdus Salam, and Steven Weinberg avaient permis dans les années 1960 d’obtenir une description unifiée à haute énergie (246 GeV, soit $246 \times 10^{9}$ eV, 1 eV correspondant à $1,6\times 10^{-19}$ joules) de l'interaction électromagnétique et de l'interaction faible en une interaction unique appelée électro-faible\footcite{Glashow1961}\footcite{Weinberg1967}\footcite{Salam1968}\footcite{Glashow1980}\footcite{Weinberg1980}\footcite{Salam1980}. Cette unification fut une brique essentielle du ``modèle standard de la physique des particules" dont la dernière brique a été détectée directement en 2012 au Large Hadron Collider (LHC) près de Genève : le boson de Brout-Englert-Higgs\footcite{Aad2012PhLB..716....1A}. L'existence hypothétique de cette particule introduite au début des années 1970 était une des clefs de voûte du modèle standard de la physique des particules, et sa détection au LHC lui a permis de passer du statut d'hypothèse mathématique fructueuse à une forme de réalité physique tangible, le terme ``réalité" étant ici entendu dans le sens de ``validée par une expérience directe et reproductible". Idéalement, c'est ce changement de statut par lequel devrait aussi passer l'hypothèse de la matière noire.

Dans la continuité de l'unification électro-faible, les physiciens envisagèrent une unification plus large des interactions fondamentales, dite ``supersymétrique", dans laquelle plus rien ne distinguerait les fermions des bosons -- les deux types de particules élementaires dans le cadre du modèle standard de la physique des particules -- au-delà d'énergies de l'ordre du TeV ($10^{12}$ eV). Une conséquence importante de cette unification serait d’associer aux particules fermioniques ordinaires des particules bosoniques ``superpartenaires". La moins massive de ces dernières pourrait être stable et constituer un candidat idéal pour la matière noire ! Ces hypothétiques particules massives et interagissant via l'interaction faible sont appelées les ``Weakly Interacting Massive Particles", ou WIMPs\footcite{Steigman1985}\footcite{Jungman1996PhR...267..195J}. Avec cette hypothèse, on pouvait faire d’une pierre deux coups, en proposant via une unification fondamentale une solution toute faite au problème de la matière noire. 

Mais la beauté de cette hypothèse ne s'arrêtait pas là. Pour calculer la densité attendue de matière noire dans l'Univers (26\% de la densité critique) on peut supposer que celle-ci est initialement dans un bain à l'équilibre thermique, où production et annihilation de matière noire s'équilibrent et dont la température diminue avec l'expansion de l'Univers. Lorsque cette température descend en-dessous de l'énergie associée à la masse de la particule supersymétrique, la réaction de production s'arrête, tandis que la réaction d'annihilation continue quelque temps, jusqu'à ce que l'expansion augmente suffisamment la séparation entre les particules : à ce moment, l'abondance de matière noire se fixe ``pour l'éternité", dans un processus que l'on nomme ``freeze-out". Dans ce processus, la densité finale prédite ne dépend que de la masse des particules et de leur section efficace d'auto-annihilation. Si l'on multiplie la vitesse correspondant à l'énergie de masse au repos typiquement attendue pour les WIMPs supersymétriques, entre 100~GeV et 1~TeV, par la section efficace (exprimée en unités de surface) associée à l'interaction électro-faible, on obtient $3 \times 10^{-26} {\rm cm}^3/{\rm s}$, et ceci donne presque précisément la bonne densité de matière noire ! Cela fut appelé le  ``miracle du WIMP''.

Avec l'hypothèse des WIMPs, tout semblait donc en place pour une résolution rapide et particulièrement élégante du problème de la matière noire, et ce d'autant plus que cela provenait de considérations théoriques indépendantes du problème lui-même. Trois types d'expérience furent mis en place pour détecter ces particules. Tout d'abord, des WIMPs auraient pu être produites dans les désintégrations liées aux collisions de particules au LHC. Ensuite, leur auto-annihilation aurait pu produire des rayons cosmiques à haute énergie depuis les régions des galaxies où une forte densité de matière noire était attendue. Enfin, des collisions directes entre particules de matière noire et noyaux de matière baryonique auraient pu être détectées dans des expériences spécifiquement mises en place à cet effet. Las ! Strictement aucune production ou détection de telles particules n'a eu lieu à ce jour\footcite{Aalbers2022arXiv220703764A}, et les expériences mises en place permettent d'exclure des portions de plus en plus larges de l'espace des paramètres en termes de masse, section efficace d'auto-annihilation ou section efficace d'interaction avec la matière baryonique. Tout l'espace des paramètres n'est pas encore exploré à ce jour, et l'on ne peut exclure qu'une détection de WIMPs ait lieu dans les prochaines années, mais le ``miracle du WIMP", lui, est mort et enterré, ce qui diminue largement l'attrait théorique de cette solution au problème de la matière noire.

Ainsi plongés dans l'abîme de notre ignorance, le point positif de cette non-détection est qu'elle ouvre l'espace des possibles, un espace particulièrement grand en l'absence de motivations théoriques fermement établies pour le restreindre. De nombreux autres candidats que les WIMPs sont ainsi explorés aujourd'hui, comme les axions --~des particules hypothétiques supposées stables, neutres et de faible masse, initialement motivées par des considérations portant sur l'interaction forte~--, 
les neutrinos stériles --~des neutrinos hypothétiques motivés par la masse non-nulle des neutrinos ordinaires~--, ou même des trous noirs primordiaux qui auraient été créés à la fin de l'ère d'inflation. Chacune de ces hypothèses nécessite de mettre en place des protocoles observationnels et expérimentaux permettant de les tester, mais le modèle cosmologique lui-même fait face à un certain nombre de défis observationnels qui pourrait remettre en cause sa validité. Même si celui-ci rend compte avec une précision impressionnante des observations de l'Univers à grande échelle, le diable se cache parfois dans les détails. 

\section{Les défis pour le modèle cosmologique actuel}



Le modèle $\Lambda$CDM permet de rendre compte avec précision du fond diffus cosmologique et de décrire l'Univers à grande échelle. Les simulations cosmologiques effectuées dans le cadre de ce modèle produisent en outre des univers simulés qui ressemblent à bien des égards à l'Univers observé, permettant de suivre la formation et l'évolution des structures comme la toile cosmique, les amas de galaxies et les galaxies elles-mêmes en partant de la grande homogénéité du fond diffus cosmologique et reproduisant une grande partie des propriétés observées. Cependant, certaines de ces propriétés posent encore des problèmes observationnels au modèle $\Lambda$CDM et pourraient en indiquer les limites. 


On observe notamment une très forte corrélation, connue sous le nom de relation de Tully-Fisher baryonique\footcite{Tully1977A&A....54..661T}\footcite{McGaugh2000ApJ...533L..99M}, entre la masse baryonique et la vitesse de rotation asymptotique dans toutes les galaxies spirales et les galaxies naines en rotation, la vitesse de rotation étant fondamentalement liée au champ de gravitation. Cette corrélation était déjà explicitée plus haut dans l'équation (3) où l'on voit que la vitesse de rotation à la puissance 4 est proportionnelle à la masse {\it baryonique} des galaxies. Or, dans le cadre du modèle $\Lambda$CDM, l'on s'attendrait plutôt à une corrélation entre le champ gravitationnel et la distribution de matière \textit{totale}, dominée par la matière noire. Le champ gravitationnel qui fixe la vitesse de rotation est en effet déterminé par l'ensemble de la masse présente. Le modèle $\Lambda$CDM requiert donc à la quantité de matière baryonique d'une galaxie d'être très précisément fixée par la quantité de matière noire du halo qui l'entoure, alors même que le halo s'étend bien au-delà de la galaxie et que les processus de formation et d'évolution des galaxies comprennent des fusions violentes et aléatoires qui devraient apporter une certaine variabilité à la masse des galaxies et des halos, et donc aussi au rapport des deux. Or la dispersion intrinsèque de cette relation entre masse baryonique et vitesse de rotation est particulièrement petite, potentiellement trop petite pour être parfaitement reproduite par nos simulations actuelles\footcite{Lelli2016ApJ...816L..14L}.
Une telle corrélation entre la distribution de matière baryonique et le champ gravitationnel semble de plus être à l'\oe uvre aussi à plus petite échelle dans les courbes de rotation des galaxies spirales, lorsqu'on trace la vitesse de rotation en fonction de la distance au centre de la galaxie\footcite{Lelli2017ApJ...836..152L}\footcite{Sancisi2004IAUS..220..233S}. Chaque excès local de matière baryonique semble en effet se répercuter sur la vitesse de rotation, alors que c'est la matière noire qui domine. Ici encore, il semblerait donc y avoir un couplage étroit entre matière baryonique et matière noire, à une échelle où les simulations indiquent plutôt un découplage des deux composantes. 
Qu'il s'agisse des galaxies dans leur ensemble ou des inhomogénéités locales de la matière baryonique, un tel couplage constitue un défi pour le modèle $\Lambda$CDM.

Les simulations cosmologiques de matière noire seule prédisent une forme universelle pour les halos de matière noire\footcite{Navarro1997ApJ...490..493N}, dont la densité augmente abruptement près du centre et qui ne dépend ni de la masse des halos ni de leur histoire : le \textit{cuspide}. L'observation des courbes de rotation des galaxies naines permet de déduire la répartition de leur matière noire et indique par contre dans de nombreuses galaxies une densité quasi-constante près du centre, qu'on décrit par le terme de \textit{c\oe ur}\footcite{Oh2011AJ....141..193O}. Si cette différence parut initialement constituer un problème en soi (le \textit{problème des cuspides}), on se rendit rapidement compte que l'introduction des baryons dans les simulations permettait de reproduire des c\oe urs comme ceux qui étaient observés\footcite{Governato2012MNRAS.422.1231G}. Contrairement à la matière noire, les baryons peuvent se refroidir, former des étoiles, et sont à l'origine d'importants phénomènes dits de \textit{feedback}, terme qui regroupe différents phénomènes baryoniques qui réchauffent et mettent en mouvement les baryons eux-mêmes : vents stellaires et champs de radiation émis au cours de la vie des étoiles, explosions de supernovae, et l'action parfois violente des trous noirs supermassifs qui se trouvent au centre des galaxies. Ces phénomènes de feedback sont à l'origine d'importants mouvements de gaz dans et autour des galaxies qui peuvent affecter gravitationnellement la matière noire et aboutir à la transformation d'un cuspide en c\oe ur\footcite{Pontzen2012MNRAS.421.3464P}\footcite{El-Zant2016MNRAS.461.1745E}\footcite{Freundlich2020MNRAS.491.4523F}. Les simulations prenant en compte ces phénomènes prédisent une répartition de la matière noire entre cuspide et c\oe ur différant en fonction de la masse du halo et de l'intensité du feedback. 
Mais plusieurs problèmes demeurent : non seulement les simulations ne s'accordent pas toutes sur l'intensité de ces phénomènes et leur effet sur la répartition de la matière noire, mais comme ces phénomènes sont {\it a priori} les mêmes d'une galaxie à l'autre, un halo de masse donnée devrait toujours avoir la même répartition de matière noire, c'est-à-dire engendrer la même courbe de rotation, or on observe une grande diversité des courbes de rotation à masse totale donnée\footcite{Oman2015MNRAS.452.3650O}\footcite{Bullock2017ARA&A..55..343B}\footcite{Ghari2019A&A...623A.123G}. 
Cela pourrait bien sûr s'expliquer par des effets de feedback différents d'une galaxie à l'autre. Mais alors, comment expliquer l'uniformité des courbes de rotation en fonction de la matière ordinaire, tant localement que globalement ? Cela nécessite un ajustement fin du feedback, qui n'est pas encore compris aujourd'hui dans le cadre du modèle $\Lambda$CDM.

Par ailleurs, des dizaines de petites galaxies satellites orbitent autour de la Voie Lactée, de la galaxie d'Andromède et des autres galaxies. Ces galaxies naines ont constitué et continuent à constituer un défi pour le modèle $\Lambda$CDM\footcite{Sales2022NatAs...6..897S}.
%
Le faible nombre de galaxies satellites sembla tout d'abord poser problème comparé aux simulations de matière noire seule, qui en prédisaient beaucoup plus (il s'agit du \textit{problème des satellites manquants})\footcite{Klypin1999ApJ...522...82K}\footcite{Moore1999ApJ...524L..19M}. Cependant, non seulement les observations ne permettaient pas de détecter les galaxies les moins lumineuses, mais encore l'augmentation de la résolution des simulations et l'introduction des phénomènes baryoniques permirent ensuite de réconcilier observations et théorie : il n'y avait plus de satellites manquants. 
%
On réalisa cependant que la densité de matière noire des galaxies satellites observées était plus faible que dans les simulations, ce qui laissait penser qu'une partie des halos de matière noire les plus massifs n'avaient pas formé d'étoiles alors qu'ils auraient dû le faire (il s'agit du \textit{problème too big to fail}, c'est-à-dire des galaxies trop grosses pour échouer à former des étoiles)\footcite{Boylan-Kolchin2011MNRAS.415L..40B}. Ce problème disparait toutefois si la masse du halo de la Voie Lactée est plus faible qu'estimée actuellement, ce qui n'est pas exclu par les observations, et surtout si les phénomènes baryoniques de feedback impliquent des c\oe urs de matière noire, dont la plus faible densité au centre rend les galaxies satellites plus enclines à perdre leur gaz et donc à former moins d'étoiles. 
%
%
Les observations semblent par contre indiquer que les galaxies satellites de la Voie Lactée, de la galaxie d'Andromède, de la galaxie elliptique Centaurus A ainsi que d'autres galaxies de l'Univers local évoluent préférentiellement dans des plans en rotation, ce qui demeure un problème pour le modèle $\Lambda$CDM (c'est le \textit{problème des plans de satellites})\footcite{Kroupa2005A&A...431..517K}\footcite{Pawlowski2018MPLA...3330004P}. En effet, les simulations cosmologiques $\Lambda$CDM indiquent plutôt des distributions et des mouvements beaucoup plus aléatoires pour les galaxies satellites, notamment du fait des fusions successives par lesquelles s'assemblent en partie les galaxies, et les alignements comme ceux qui sont observés y sont rares. Contrairement à la densité de la matière noire et à la formation des étoiles, les positions et les mouvements des satellites sont largement indépendants des phénomènes baryoniques et du type de matière noire. L'accrétion le long des filaments de la toile cosmique ou une accrétion par groupes peuvent être envisagées pour expliquer les plans observés, mais ces deux modes d'accrétion ne semblent pas suffisants pour expliquer la fraction des satellites concernés et la faible épaisseur des plans ; de plus, ils sont déjà pris en compte dans les simulations. Une autre possibilité envisagée est celle de galaxies naines formées par effet de marée lors d'une interaction passée. Cela impliquerait que les satellites aient été formés à partir de matériel enrichi en éléments lourds et qu'ils contiennent peu de matière noire, ce qui ne semble pas être le cas. Tout comme la diversité des courbes de rotation des galaxies naines, les alignements de satellites restent un défi pour le modèle $\Lambda$CDM, même s'ils pourraient n'être que transitoires.

Une grande partie des galaxies spirales de l'Univers local arborent une barre, c'est-à-dire une bande d'étoiles qui traverse le centre de la galaxie et qui tourne de manière cohérente. La présence d'une barre jouerait un rôle important dans l'évolution de ces galaxies, en remuant les étoiles et le gaz. La formation d'une barre serait naturelle dans un disque gravitationnellement instable, et sa persistance liée aux interactions avec le reste de la distribution de matière, en particulier le halo de matière noire dans le cadre du modèle $\Lambda$CDM. Mais les simulations cosmologiques dans ce cadre ont du mal à reproduire la fraction observée de barres et produisent généralement des barres qui tournent trop lentement ou sont trop petites par rapport aux observations\footcite{Algorry2017MNRAS.469.1054A}\footcite{Peschken2019MNRAS.483.2721P}\footcite{Fragkoudi2021A&A...650L..16F}\footcite{Roshan2021MNRAS.508..926R}\footcite{Reddish2022MNRAS.512..160R}\footcite{Frankel2022ApJ...940...61F}. La vitesse de rotation de la barre est intimement liée à la quantité de matière noire au centre de la galaxie, dans la mesure où le halo de matière noire ralentit la barre du fait de la friction dynamique\footcite{Weinberg1985MNRAS.213..451W}. Ainsi, les vitesses observées semblent indiquer que la dynamique des galaxies spirales massives est dominée par les baryons et non par la matière noire, ce qui pourrait poser un problème au modèle $\Lambda$CDM. A contrario, des mesures du ralentissement de la barre de la Voie Lactée pourraient permettre de mettre en évidence la présence de matière noire\footcite{Chiba2021MNRAS.500.4710C}.

Ainsi, de nombreux défis se posent encore au modèle cosmologique standard aux échelles galactiques. Ces défis pourraient trouver une solution dans une meilleure compréhension des mécanismes de formation et d'évolution des galaxies, mais l'on ne peut pas exclure que le problème soit en fait de nature plus fondamentale. Cette possibilité est renforcée par le fait que certaines tensions avec le modèle cosmologique $\Lambda$CDM ont également vu le jour à plus grande échelle. En particulier, on peut déduire des observations de l'Univers à grande échelle la valeur de la constante de Hubble-Lema\^itre en tant que paramètre du modèle $\Lambda$CDM, et ce avec une grande précision, $H_0 = 67,4 \pm 0,5$ (km/s)/Mpc\footcite{Planck2020A&A...641A...6P}. Or, il s'agit d'un paramètre que l'on peut également mesurer de façon directe, en utilisant la corrélation entre distance et vitesse d'expansion comme Hubble l'avait fait originellement. Dans ce cas, la valeur estimée est $H_0 = 73,04 \pm 1,04$ (km/s)/Mpc\footcite{Riess2022ApJ...934L...7R}. La différence peut paraître faible, mais elle ne l'est pas au regard des incertitudes estimées. Il s'agit là d'un grand mystère de la cosmologie moderne, pour lequel il n'y a pas de solution simple. En outre, l'amplitude des fluctuations dans le fond diffus cosmologique peut permettre de prédire à l'aide du modèle $\Lambda$CDM l'amplitude des fluctuations des structures dans l'univers tardif, comme les amas de galaxies, mais une tension semble apparaître : l'amplitude des fluctuations mesurée directement à l'échelle des amas est moins importante que prédite\footcite{Heymans2021A&A...646A.140H}. La plupart des tentatives de solution à la tension sur la valeur de la constante de Hubble-Lema\^itre ont tendance à aggraver ce problème, ou inversement, toutes les solutions éventuelles à ce problème semblent aggraver la tension de Hubble. Le mystère reste donc entier. Est-ce simplement un problème de biais systématiques dans les données elles-mêmes, ou cela nous révèle-t-il les limites du modèle cosmologique ?

Enfin, il est utile de rappeler qu'outre la matière noire, l'origine, la valeur et la nature de la constante cosmologique (ou énergie noire) reste un autre casse-tête pour la cosmologie moderne\footcite{Carroll2001LRR.....4....1C}. Comme nous allons le voir, la valeur de la constante cosmologique se retrouve aussi de façon un peu impromptue dans la dynamique des galaxies. Il pourrait s'agir d'une coïncidence, mais aussi d'un indice que quelque chose de fondamental nous échappe peut-être encore.

\section{Alternatives}

Nombre des défis pour le modèle cosmologique concernent les échelles galactiques. Cela n'est pas forcément surprenant car ces échelles sont hautement non-linéaires et nécessitent une bonne compréhension de tous les phénomènes complexes affectant la physique des baryons, en particulier les phénomènes de feedback qui peuvent influencer la répartition de la matière noire. Néanmoins, la plupart de ces mystères concernent des régularités observées là où l'on attendrait de la diversité, et de la diversité là où l'on attendrait {\it a priori} de la régularité. 

La très forte corrélation entre la matière baryonique et le champ gravitationnel dans les galaxies correspond à ce que l'on pourrait attendre si la matière baryonique {\it seule} dictait le comportement du champ gravitationnel. Dans ce cas, une modification de la gravitation pourrait rendre compte des observations à la place de la matière noire. Milgrom a proposé dès 1983 une telle modification de la gravitation (ou du principe fondamental de la dynamique, la deuxième ``loi'' de Newton) dans le régime des très faibles accélérations caractéristiques des régions externes des galaxies\footcite{Milgrom1983ApJ...270..365M}\footcite{Milgrom1983ApJ...270..371M}\footcite{Milgrom1983ApJ...270..384M}, qui sont de l'ordre de cent milliard de fois plus faibles qu'à la surface de la Terre. L'idée, connue sous le nom de ``{\it Modified Newtonian Dynamics}" ou MOND, est que lorsque l'accélération devient plus faible qu'une certaine accélération critique $a_0$ (qui serait une nouvelle constante de la nature), le mouvement cesse d'obéir à la loi de la gravitation universelle (ou, au choix, au principe fondamental de la dynamique) de Newton en tant que limite de champ faible de la relativité générale. L'accélération $g$ décroîtrait dans ce régime comme l'inverse de la distance plutôt que comme l'inverse du carré de la distance, et serait reliée à l'accélération Newtonienne $g_N$ par
\begin{equation}
\left\{
    \begin{array}{ll}
       g=g_N                &  {\rm quand} \; g \gg a_0\\
       {\rm et}\\
       g=\sqrt{g_N a_0}     &  {\rm quand} \; g \ll a_0.
    \end{array}
\right.
\end{equation}
Cela permet notamment de prédire l'équation (3). Mais il est à noter que, si une relation observationnelle existait déjà entre luminosité des galaxies et vitesse de rotation de celles-ci\footcite{Tully1977A&A....54..661T}, et que celle-ci inspira la proposition de Milgrom, il n'était pas donné que cette relation resterait vraie dans des galaxies dont la masse baryonique est largement dominée par le gaz. Or, on se rendit compte par la suite que cette prédiction de Milgrom était correcte\footcite{McGaugh2000ApJ...533L..99M}, et ce n'est pas la seule prédiction qu'il fit en 1983 quant à la dynamique des galaxies : l'idée simple qu'il proposa permettait de déduire un certain nombre d'analogues aux lois empiriques de Kepler pour les planètes du Système Solaire, mais cette fois à l'échelle des galaxies\footcite{Famaey2012LRR....15...10F}. 
Par exemple, Milgrom prédit explicitement en 1983 une corrélation entre la valeur de la densité de surface d'une galaxie et la pente avec laquelle la vitesse de rotation augmente jusqu'à sa valeur asymptotique. Les galaxies à faible brillance de surface permettant de tester cette hypothèse n'ont été découvertes qu'une décennie plus tard : il s'agissait donc d'une prédiction  {\it a priori} de la diversité des formes des courbes de rotation que l'on observe aujourd'hui, vérifiée {\it a posteriori} par l'observation. 
Interprétée en termes de ``matière noire", cette prédiction signifie que pour un ``halo" de masse donnée (qui engendre le même champ gravitationnel que la modification de la gravitation proposée par Milgrom), la courbe de rotation peut avoir des formes très différentes, avec un ``cœur" de densité quasi-constante ou avec un ``cuspide", en fonction de la densité de surface de la matière {\it baryonique}. Or, c'est exactement ce qui est observé ! Et c'est précisément ce type d'observations qui pose des problèmes d'ajustement fin dans le cadre $\Lambda$CDM standard.

De plus, les disques galactiques sont plus instables sans matière noire et forment plus facilement des barres, tandis que celles-ci ne sont pas ralenties par les halos de matière noire\footcite{Tiret2007A&A...464..517T}\footcite{Roshan2021MNRAS.503.2833R}, en accord avec les observations mentionnées dans la section précédente. Les galaxies satellites de la Voie Lactée et d'Andromède auraient quant à elles pu naître dans une interaction passée entre ces deux galaxies, peu probable dans le cadre $\Lambda$CDM standard mais justifiée dans le cas de MOND, formant potentiellement naturellement les plans de satellites observés\footcite{Bilek2021Galax...9..100B}\footcite{Banik2022MNRAS.513..129B}.
Enfin, on peut noter que la nouvelle constante de la nature proposée par Milgrom, l'accélération caractéristique $a_0 \sim 10^{-10} ~{\rm m}/{\rm s}^2$ est en fait de l'ordre de la racine carrée de la constante cosmologique $\Lambda$ (en unités ``naturelles", où la vitesse est exprimée en multiples de la vitesse de la lumière), montrant qu'il existerait peut-être un lien entre l'énergie noire et les effets que l'on attribue à la matière noire ! Milgrom a lui-même proposé un tel lien en 1999\footcite{Milgrom1999PhLA..253..273M} : dans un espace-temps plat (dit de Minkowski), en l'absence de courbure liée à la présence de masse, un observateur accéléré voit théoriquement le vide comme un bain thermique dont la température est proportionnelle à l'accélération de l'observateur. C'est ce que l'on appelle le rayonnement de Unruh\footcite{Unruh1976PhRvD..14..870U}, et cela signifie que l'accélération peut être redéfinie comme étant proportionnelle à la température du rayonnement de Unruh. D'autre part, un observateur accéléré dans un univers affecté par une constante cosmologique positive (univers dit de ``de Sitter") voit quant à lui une combinaison non linéaire du rayonnement du vide de Unruh et du rayonnement de Gibbons-Hawking dû à l'horizon cosmologique\footcite{Gibbons1977PhRvD..15.2738G}. Si l'on remplace, dans le principe fondamental de la dynamique de Newton, l'accélération d'un observateur non-inertiel par la différence de température entre celle qu'il voit et celle qu'il verrait en étant ramené dans un état inertiel (c'est-à-dire dans lequel il ne verrait que le rayonnement de Gibbons-Hawking), on obtient précisément la relation proposée par Milgrom en 1983, avec une accélération caractéristique de l'ordre de la racine carrée de la constante cosmologique ! Les observateurs subissant une très faible accélération verraient un rayonnement de Unruh avec une basse température, proche de celle de Gibbons-Hawking, et verraient donc leur dynamique modifiée, tandis que les observateurs subissant une accélération plus grande ne verraient pas de différence notable avec une température de Gibbons-Hawking nulle (comme en l'absence de constante cosmologique). L'accélération serait donc plus grande dans un Univers avec constante cosmologique positive que dans la dynamique Newtonienne, et ce en produisant précisément la relation hypothétisée par Milgrom en 1983.  Malheureusement, aucune modification de la relativité générale basée sur cette approche impliquant une modification du principe fondamental de la dynamique n'a encore été rigoureusement développée, si tant est que cela soit même possible. En 2017, le physicien néerlandais Erik Verlinde a repris ces idées de Milgrom dans le cadre d'une gravitation dite ``émergente" dans un espace-temps de de Sitter\footcite{Verlinde2017ScPP....2...16V}, mais cette théorie n'est pas non plus rigoureusement formalisée à ce jour.

La proposition de Milgrom peut néanmoins aussi être formalisée non par une modification du principe fondamental de la dynamique mais par une modification non-linéaire de la gravitation Newtonienne en champ faible, où l'accélération caractéristique $a_0$ est alors simplement introduite en tant que nouvelle constante de la nature indépendamment de la constante cosmologique. Ce type de formulation rend compte, sans matière noire, de la plupart des phénomènes observés aux échelles galactiques, de façon beaucoup plus naturelle que le modèle $\Lambda$CDM. Mais pourquoi, alors, n'est-ce pas la solution universellement acceptée au problème de la matière noire, qui s'apparenterait dès lors à l'explication qui fut trouvée pour expliquer l'anomalie de l'orbite de Mercure (un nouveau modèle théorique) plutôt qu'à l'explication trouvée pour celle d'Uranus (la présence d'une masse invisible) ?

La réponse à cette question est que l'application de la modification de la gravitation Newtonienne proposée par Milgrom aux plus grandes échelles (celle des amas de galaxies et au-delà) et aux plus petites (aux échelles inférieures au parsec, comme celles du Système Solaire) pose plus de problèmes qu'elle n'apporte de solutions\footcite{Famaey2012LRR....15...10F}. Tout d'abord, dans les amas de galaxies, la modification proposée ne rend pas compte de la totalité de la masse manquante, car l'accélération est typiquement supérieure à l'accélération critique dans les parties centrales alors que le besoin de matière noire s'y fait quand même sentir. Il faut donc de nouveau recourir à de la matière noire. De plus, l'observation en 2006 de l'effet de lentillage gravitationnel de l'amas dit ``du boulet", qui résulte de la collision de deux amas de galaxies, a montré que cette matière noire devrait se comporter comme on l'attend de la matière noire froide dans le cadre standard, c'est-à-dire sans ralentissement lié à la collision\footcite{Clowe2006ApJ...648L.109C}. Par ailleurs, la dynamique interne de certaines galaxies ``ultra-diffuses" dans l'amas de Coma semble bien obéir aux prédictions de MOND, mais elles le font comme si elles étaient isolées de l'amas, ce qui ne devrait pas être le cas\footcite{Freundlich2022A&A...658A..26F}. Enfin, comme nous l'avons vu précédemment, la matière noire non-baryonique est un élément essentiel à notre compréhension de la structure détaillée du fond diffus cosmologique et de la formation des structures. S'il n'y a pas de matière noire à ces échelles, il doit quand même y avoir quelque chose qui se comporte ``comme" de la matière noire. Finalement, dans le Système Solaire, les effets attribués à la modification de la gravitation seraient faibles mais pourraient en principe être détectables, or rien de tel n'a été détecté à l'heure actuelle\footcite{Blanchet2011MNRAS.412.2530B}\footcite{Hees2016MNRAS.455..449H} : cela n'exclut pas une modification mais oblige à un ajustement fin de celle-ci à ces échelles, ce qui réduit potentiellement son attrait.

Pour obtenir la modification non-linéaire de la gravitation Newtonienne en champ faible proposée par Milgrom, il faut modifier la relativité générale de manière cohérente. Ceci peut être fait en ajoutant des champs à celle-ci, en plus du champ ``métrique" décrivant la courbure de l'espace-temps. C'est le physicien israélien Jacob Bekenstein, père de la théorie de l'entropie des trous noirs, qui proposa pour la première fois en 2004 une telle modification relativiste, en ajoutant à la métrique un champ de quadrivecteurs et un champ scalaire\footcite{Bekenstein2004PhRvD..70h3509B}. Cette modification a été affinée et généralisée depuis lors, et sa dernière version a été proposée en 2021 par Constantinos Skordis et Tom Zlo\'snik\footcite{Skordis2021PhRvL.127p1302S}. Cette théorie reproduit la gravitation ``MONDienne" à l'échelle des galaxies, mais les champs additionnels peuvent également se comporter comme de la matière noire aux échelles cosmologiques, permettant ainsi de reproduire le fond diffus. Il reste encore à voir si elle peut également expliquer la dynamique des amas de galaxies et passer les contraintes aux échelles inférieures au parsec. 

Quoi qu'il en soit, l'ajout de ces nouveaux champs pouvant se comporter comme de la matière noire aux échelles cosmologiques mais engendrant une modification de la gravitation en régime de champ faible quasi-statique (le régime des galaxies) nous indique que le débat entre matière noire et gravitation modifiée a une teneur qui pourrait être partiellement de nature sémantique. Si ces champs se comportent comme de la matière noire aux échelles cosmologiques, on peut être en droit de les appeler ``matière noire", si ce n'est qu'ils se comportent très différemment de la matière noire froide du modèle $\Lambda$CDM aux échelles galactiques. Entre ces deux extrêmes, il existe toute une série de possibilités intermédiaires de modifications du modèle $\Lambda$CDM changeant les propriétés de la matière noire au sein de celui-ci. Parmi ces modèles, certains sont spécifiquement pensés de façon à reproduire la phénoménologie de MOND dans les galaxies, comme le modèle de matière noire appelée ``superfluide"\footcite{Berezhiani2015PhRvD..92j3510B}\footcite{Berezhiani2018JCAP...09..021B} ou encore le modèle de matière noire ``dipolaire"\footcite{Blanchet2008PhRvD..78b4031B}\footcite{Stahl2022MNRAS.517..498S}. D'autres modèles proposent des modifications plus mineures de la nature fondamentale de la matière noire, qui ne reproduisent pas {\it stricto sensu} la phénoménologie de MOND, mais qui peuvent potentiellement alléger certaines tensions, notamment sur la répartition entre c\oe urs et cuspides de matière noire. 
C'est le cas de modèles tels que ceux de matière noire ``floue"\footcite{Hu2000PhRvL..85.1158H}\footcite{Schive2014NatPh..10..496S}\footcite{Hui2017PhRvD..95d3541H} ou encore de matière noire auto-interagissante\footcite{Spergel2000PhRvL..84.3760S}\footcite{Peter2013MNRAS.430..105P}\footcite{Elbert2015MNRAS.453...29E}. 
Le modèle de matière noire floue suppose que les particules de matière noire sont tellement légères (avec des masses de l'ordre de $10^{-22}~\rm eV$, soit 33 ordres de grandeur de moins que les WIMPSs) qu'elles induisent des effets quantiques à l'échelle des galaxies : franges d'interférences, fluctuations de densité, et présence d'un c\oe ur de matière noire. De tels effets pourraient affecter les structures stellaires\footcite{Marsh2019PhRvL.123e1103M}\footcite{El-Zant2020MNRAS.492..877E}, dont l'observation ou la non-observation pourrait permettre de valider ou d'infirmer la théorie. 
Le modèle de matière noire auto-interagissante, quant à lui, suppose que les particules de matière noire puissent interagir entre elles. Ce modèle est très proche du modèle $\Lambda$CDM aux échelles cosmologiques mais s'en détache à l'échelle des galaxies, dans la mesure où la plus forte densité favorise les interactions. La matière noire auto-interagissante induit notamment des densités de matière noire moins importantes au centre des halos que la matière noire froide. 
Tous ces modèles nécessitent la mise en place de simulations aussi poussées que celles réalisées dans le cadre $\Lambda$CDM standard pour en explorer les prédictions, les bienfaits éventuels ainsi que les limites. Enfin, les alternatives à la constante cosmologique sont également envisagées, notamment via des modifications de la gravitation\footcite{Clifton2012PhR...513....1C}, et seront testées par les prochaines générations de missions spatiales comme la mission Euclid.



\section{Conclusion}

Dans cet essai, nous avons entrepris de donner une brève vue d'ensemble,\linebreak forcément non-exhaustive, de l'histoire des idées qui ont collectivement mené à l'établissement d'un modèle standard de la cosmologie connu sous le nom de modèle $\Lambda$CDM. Ce modèle repose 
sur l'existence de deux composantes ``sombres", qui à notre époque dans l'évolution de l'Univers représentent respectivement 69\% du bilan énergétique pour l'{\it énergie noire} et 26\% pour la {\it matière noire}, les 5\% restant étant la matière ordinaire, aussi appelée baryonique. Ce modèle rend compte avec une précision impressionnante des observations de l'Univers à grande échelle, et présente une grande résilience face aux nouvelles observations à ces échelles, à l'exception de deux tensions notables : la valeur de la constante de Hubble-Lema\^itre qui intervient dans le modèle ainsi que l'amplitude des fluctuations des structures dans l'Univers tardif. De plus, les particules de matière noire n'ont toujours pas été détectées de manière non-gravitationnelle, malgré les efforts des grands collisionneurs de particules comme le LHC. 

Un certain nombre de coïncidences frappent par ailleurs l'attention. En particulier, on remarque que les trois composantes principales du bilan énegétique de l'Univers sont toutes d'un ordre de grandeur similaire. Dans le cas de la matière noire et de la matière baryonique, représentant chacune respectivement 84\% et 16\% du contenu en matière de l'Univers, cet ordre de grandeur devrait être le même depuis l'époque de la génèse des baryons. Pourtant rien, dans les mécanismes habituellement invoqués pour fixer la densité des baryons et celle de la matière noire, ne devrait impliquer que les densités respectives ne soient pas différentes de plusieurs ordres de grandeur. La coïncidence liée à la densité d'énergie noire actuelle, qui est également d'un ordre de grandeur similaire à celles des deux autres composantes, renforce le sentiment qu'un lien entre ces différentes composantes nous échappe peut-être. Mais surtout, d'autres coïncidences inquiétantes se font jour aux échelles galactiques : les phénomènes attribués à la matière noire y semblent en effet intimement liés à la distribution de la matière baryonique, et la phénoménologie observée peut être résumée par une relation assez simple faisant intervenir une accélération caractéristique de l'ordre de la racine carrée de la constante cosmologique. Cette relation entre champ gravitationnel engendré par la matière noire dans les galaxies et distribution de la matière baryonique avait été prédite par Milgrom en 1983. La modification de la dynamique Newtonienne qu'il proposa à cette époque, connue sous le nom de MOND, rend compte de façon remarquable de nombreuses observations aux échelles galactiques. Malheureusement, l'application d'une telle modification aux échelles extra-galactiques se heurte à un certain nombre de problèmes difficiles à surmonter, et ce précisément à des échelles où les observations sont en excellent accord avec le modèle $\Lambda$CDM. À l'échelle du Système Solaire, aucune indication d'une modification de type MOND n'a jamais été détectée. 

Face à ces différents constats, certaines interrogations qui ont marqué l'histoire des sciences et en particulier de l'astronomie semblent être toujours d'actualité et pouvoir nous éclairer. Quel est le statut ontologique des hypothèses avancées par les modèles ? Faut-il y voir uniquement des fictions mathématiques ou des réalités concrètes ? Pouvons-nous atteindre quelque chose de la nature des choses ou devons-nous nous contenter de sauver les phénomènes ? Comment différencier deux modèles très différents quand ils rendent compte des phénomènes tous deux à leur manière et dans des domaines différents ? Et plus précisément concernant la matière noire et la gravitation modifiée : l'ajout de champs en gravité modifiée rend-elle la question purement sémantique ? Faut-il envisager des formes ``intermédiaires" de matière noire ``modifiée", comme la matière noire floue et la matière noire auto-interagissante, voire toute une zoologie de particules sombres ? MOND pourrait-elle être une loi empirique émergeant des propriétés de la matière noire et de son interaction avec les baryons ? La matière noire pourrait elle n'être qu'une paramétrisation d'une modification plus fondamentale de la gravitation ? Ou MOND et matière noire sont-elles toutes deux des hypothèses vouées à être supplantées par un changement de point de vue digne de la révolution Copernicienne, sous-tendue par des hypothèses moins intuitives que les notions de force, d'espace-temps et de particule auxquelles nous avons été habitués ? Car s'il y a bien une chose que l'histoire de la physique nous a montré, c'est que la nature est toujours plus complexe que la grille de lecture que nous en avons à un instant donné. 

\newpage


\printbibliography

\end{document}